# A Random Process Model Useful for Describing Radar Clutter

Dr. K. James Sangston
Sensors and Electromagnetic Applications Laboratory
Georgia Tech Research Institute
Georgia Institute of Technology
Atlanta, GA 30332-0858
404 407-7677
jim.sangston@gtri.gatech.edu; jsangston@mac.com

**Abstract**

We use the theory of Bernstein functions, completely monotonic functions, and Levy processes to define a positive random process $\tau(t)$. For radar clutter one may think of $\tau(t)$ as the instantaneous power of the scattered radar signal that is described by a compound-Gaussian model. Thus the results herein give a mechanism for defining and simulating a compound-Gaussian random process that can be used in various radar studies. We give several examples of the sample paths of this process.

**Keywords:** radar clutter, compound-Gaussian models, Bernstein functions, completely monotonic functions, Levy process, scattered field

**Introduction**

A popular model of the statistics of a single complex radar return $z$ from the surface of the sea is the compound-Gaussian model with probability density function (PDF)

$$p(z) = \int_0^\infty \frac{e^{-\frac{|z|^2}{\sigma^2 \tau}}}{\pi \sigma^2 \tau} f(\tau) d\tau \qquad (1)$$

where $\tau$ is a unit-mean positive random variable with PDF $f(\tau)$ [1,2]. With this model, the complex random variable $z$ may be modeled as

$$z = \sqrt{\tau} x \qquad (2)$$

where $x$ is a zero-mean circular complex Gaussian random variable with variance $\sigma^2$ and $\tau$ is independent of $x$. This model has proven to be quite useful to describe the univariate statistics of radar clutter returns. The form of (2) suggests that we seek to extend this model to a random process model of the form

$$z(t) = \sqrt{\tau(t)} x(t) \qquad (3)$$

where $x(t)$ is a zero-mean complex Gaussian process with a specified covariance function and $\tau(t)$ is a non-negative random process independent of $x(t)$. The modeling of a complex Gaussian process $x(t)$ has been much-studied; consequently this work focuses on developing a model for $\tau(t)$.

To begin let $X_i \in \mathbb{R}^m, i = 1, 2, \cdots$ be a sequence of $m$-dimensional random vectors and examine the sum

$$S_N = \frac{1}{\sqrt{N}} \sum_{i=1}^{N} X_i \qquad (4)$$

where $N$ is a non-random integer. Note that the complex value $x_i = a_i e^{j\phi_i}$ may be modeled by a two-dimensional vector $X_i$, and thus this form encompasses a phenomenological model of the scattered return [3]:

$$E = \frac{1}{\sqrt{N}} \sum_{i=1}^{N} a_i e^{j\phi_i} \qquad (5)$$

From (4) assume that the central limit theorem implies

$$X = \lim_{N \to \infty} S_N \qquad (6)$$

where $X$ is an $m$-dimensional Gaussian random vector with probability density function (PDF)

$$p_X(x) = \frac{e^{-\frac{1}{2} x^T R^{-1} x}}{(2\pi)^{\frac{m}{2}} |R|^{\frac{1}{2}}} \qquad (7)$$

and $R$ is a positive definite covariance matrix. For example a single circular complex sample $x = \lim_{N \to \infty} E$ may be modeled by a two-dimensional real vector with covariance $R = \frac{\sigma^2}{2} I$.

Now modify the problem and assume $N$ is a random integer with mean $\overline{N}$ and examine the sum

$$Z_{\overline{N}} = \frac{1}{\sqrt{\overline{N}}} \sum_{i=1}^{N} X_i \qquad (8)$$

The normalization is now by $\sqrt{\overline{N}}$ rather than $\sqrt{N}$. Define a random vector $Z$ as

$$Z = \lim_{\overline{N} \to \infty} Z_{\overline{N}} \qquad (9)$$

Under quite general conditions [4-8] this random vector exists and has a compound-Gaussian PDF given by

$$p_Z(z) = \int_0^\infty \frac{e^{-\frac{1}{2\tau} z^T R^{-1} z}}{(2\pi\tau)^{\frac{m}{2}} |R|^{\frac{1}{2}}} dF_\tau(\tau) \qquad (10)$$

where $F_\tau(\tau)$ is the cumulative distribution function (CDF) of a non-negative random variable $\tau$ given by

$$\tau = \lim_{\overline{N} \to \infty} \frac{N}{\overline{N}} \qquad (11)$$

with $E[\tau] = 1$. The limit in (11) converges in distribution. The characteristic function of $Z$ then takes the form

$$C_Z(t) = G\left(\frac{t^T R t}{2}\right) \qquad (12)$$

where [9]

$$G(z) = \int_0^\infty e^{-z\tau} dF_\tau(\tau) \tag{13}$$

It is natural in this case to write the compound-Gaussian random vector $\mathbf{Z}$ as

$$\mathbf{Z} = \sqrt{\tau}\mathbf{Y} \tag{14}$$

where $\mathbf{Y}$ is a Gaussian random vector and $\tau$ is a positive random variable independent of $\mathbf{Y}$. The result in (11) now suggests that one may model the random process in (3) by developing a model for $\tau(t)$ from a number fluctuation model for the number of scatterers $N(t)$. In particular, for a given $N(t)$ one would form the limit

$$\tau(t) = \lim_{\overline{N} \to \infty} \frac{N(t)}{\overline{N}} \tag{15}$$

This is the approach followed herein.

Empirically, models for the random variable $\tau$ used to model radar clutter statistics include (among others) the gamma, inverse gamma, and generalized inverse Gaussian distributions [2,10-11]. Interestingly, each of these distributions is infinitely divisible, which suggests that one approach to modeling $\tau(t)$ would be to develop an infinitely divisible model for the number fluctuation process $N(t)$. To this end we make the following observations:

1. As discussed more fully below, the function $G(z)$ in (13) that defines the compound-Gaussian characteristic function in (12) is a completely monotonic function on the interval $z \geq 0$.
2. The random variables $\tau$ associated with several different important empirically adopted compound-Gaussian models are infinitely divisible.
3. If the random variable $\tau$ associated with a completely monotonic function $G(z)$ is infinitely divisible, then the function $H(z) = -\ln G(z)$ is a Bernstein function.
4. Every Bernstein function defines a Levy process, which is the class of processes associated with infinitely divisible random variables. (For the benefit of the reader not familiar with Levy processes we include a brief overview in the appendix.)

Hence, in cases where the univariate random variable $\tau$ is infinitely divisible – which is the case for almost all models that have been used to empirically model radar clutter – one may use a Levy process to develop a model of the number fluctuation process $N(t)$, which may in turn be used to model $\tau(t)$.

In the development that follows we start with a Bernstein function $h(z)$ and use it to define the point mass function (PMF) of a positive integer-valued random variable $K$. The random integer $K$ is then used as a mixing variable to define a non-negative integer-valued compound-Poisson process $N(t)$. The process $N(t)$ is not quite suitable for our purposes – it is non-decreasing and we want to model scenarios in which the number of scatterers can increase or decrease as time passes. To address this issue we define a windowed process $N_T(t)$ as a segment of $N(t)$ where $T$ is the length of a segment and the location of the segment varies with time. As a result, scatterers move in and out of the window and give the desired fluctuation behavior. Because the Levy process $N(t)$ has stationary and independent increments and because $N_T(t)$ at any given time is essentially an increment of $N(t)$, it follows that $N_T(t)$ is a stationary process. We then define a normalized process $\tau_{\overline{N_T}}(t) = N_T(t)/\overline{N_T}$ where $\overline{N_T}$ is the mean value of $N_T(t)$. The desired stationary process $\tau(t)$ is obtained by letting $\overline{N_T} \to \infty$. The process $\tau(t)$ has univariate statistics defined by the CDF $F_\tau$ associated with a completely monotonic function $G(z)$ that we obtain from $h(z)$.

**A Class of Infinitely Divisible Discrete Random Variables**

A function $G(z)$ is completely monotonic on the interval $z > 0$ if it has derivatives of all orders satisfying [12-15]

$$(-1)^k G^{(k)}(z) \geq 0, k = 0, 1, 2, \cdots \tag{16}$$

It is further said to be completely monotonic on the interval $z \geq 0$ if $G(0) = G(0^+) < \infty$. By the Bernstein-Widder theorem [16] a completely monotonic function on the interval $z > 0$ may be written as in (13) where the function $F_\tau(\tau)$ is non-decreasing but need not be a CDF in general. However, herein we assume

$$G(0) = \int_0^\infty dF_\tau(\tau) = 1 \tag{17}$$

With this assumption $F_\tau(\tau)$ is the CDF of a non-negative random variable $\tau$. Because we will ultimately use $G(z)$ to model a characteristic function, the assumption in (17) is a natural one. The moments of $\tau$, if they exist, are given by

$$E[\tau^m] = (-1)^m G^{(m)}(0) = \int_0^\infty \tau^m dF_\tau(\tau) \tag{18}$$

A class of functions closely related to completely monotonic functions is the class of Bernstein functions [15]. A positive function $h(z) \geq 0$ is a Bernstein function if $h^{(1)}(z)$ is completely monotonic on $z > 0$. A positive function $h(z)$ is a Bernstein function if and only if it has a Levy-Khintchine representation

$$h(z) = a + bz + \int_0^\infty (1 - e^{-zs}) dL(s) \tag{19}$$

where $a, b \geq 0$ and $L$ is a non-decreasing function on the positive real half-line such that

$$\int_0^\infty \min(1, x) \, dL(x) < \infty \tag{20}$$

In this work we assume

$$\lim_{z \to \infty} \frac{h(z)}{z} = 0 \tag{21}$$

$$h(0) = 0 \tag{22}$$

These assumptions imply

$$a = b = 0 \tag{23}$$

We further restrict $h(z)$ to satisfy

$$h^{(1)}(0) = h_1 < \infty \tag{24}$$

$$-h^{(2)}(0) = -h_2 < \infty \tag{25}$$

These latter restrictions arise from certain requirements we impose on the resulting clutter process and may not be necessary for other applications.

Let $h(z), z \geq 0$ be a Bernstein function satisfying (21-22, 24-25). Let $\kappa > 0$. It can be shown that one may use $h(z)$ to define a discrete random variable $K$ with probability generating function (PGF)

$$E[u^K] = \phi_K(u;\kappa) = 1 - \frac{h(\kappa(1-u))}{h(\kappa)}, 0 < u \leq 1 \tag{26}$$

and corresponding PMF

$$\Pr[K = n] = p_K(n;\kappa) = \begin{cases} -\frac{(-\kappa)^n}{n!} \frac{h^{(n)}(\kappa)}{h(\kappa)}, & n = 1, 2, \cdots \\ 0, & n = 0 \end{cases} \tag{27}$$

The mean value of $K$ is given by

$$E[K] = \overline{K} = \frac{\kappa}{h(\kappa)} h_1 \tag{28}$$

The condition in (21) implies that $\overline{K} \to \infty$ iff $\kappa \to \infty$, a fact that we use below.

Let $\gamma > 0$ and define

$$\lambda(\kappa) = \gamma h(\kappa) \tag{29}$$

Let $K_i, i = 1, 2, \cdots$ be independently and identically distributed (IID) random variables with PMF $p_K(n;\kappa)$ and let $P_\lambda(t)$ be a Poisson counting process with intensity $\lambda(\kappa)$. Let $T > 0$ and define a windowed compound Poisson process as

$$N_T(t) = \sum_{i=P_\lambda(t)+1}^{P_\lambda(t+T)} K_i \tag{30}$$

If $K_i$ represents a cluster of points, then the number of clusters within the window at time $t$ is a random integer given by the number of Poisson points between $t$ and $t + T$. This scenario is illustrated in Figure 1:

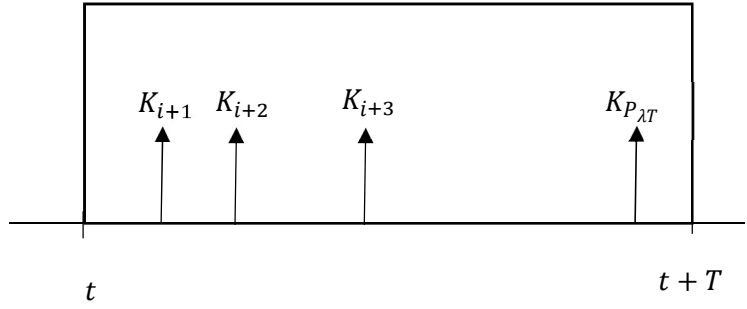

**Figure 1.** Physical Interpretation of the Windowed Compound-Poisson Process $N_T(t)$

The PGF of $N_T(t)$ is given by

$$E[u^{N_T(t)}] = \phi_{N_T(t)}(u) = e^{-\lambda T(1-\phi_K(u;\kappa))} = e^{-\gamma Th(\kappa(1-u))} \qquad (31)$$

The PGF is independent of $t$. From this result we find that the mean of $N_T(t)$ is given by

$$E[N_T(t)] = \overline{N_T} = \phi^{(1)}_{N_T(t)}(1) = \gamma h(\kappa) T\overline{K} = \lambda T\overline{K} \qquad (32)$$

Define

$$\tau_{\overline{N_T}}(t) = \frac{N_T(t)}{\overline{N_T}} \qquad (33)$$

Let

$$u = e^{-\frac{z}{\overline{N_T}}}, 0 \le z < \infty \qquad (34)$$

We now find

$$\phi_K\left(e^{-\frac{z}{\overline{N_T}}};\kappa\right) = E\left[e^{-\tau_{\overline{N_T}}(t)z}\right] = e^{-\gamma Th\left(\left(1-e^{-\frac{z}{\kappa\gamma Th_1}}\right)\right)} \qquad (35)$$

We have $\kappa \to \infty$ iff $\overline{N_T} \to \infty$. Hence

$$\lim_{\overline{N_T} \to \infty} E\left[e^{-\tau_{\overline{N_T}}(t)z}\right] = \lim_{\kappa \to \infty} e^{-\gamma T h\left(\kappa\left(1-e^{-\frac{z}{\kappa \gamma T h_1}}\right)\right)} = e^{-vh\left(\frac{z}{vh_1}\right)} \quad (36)$$

where we have defined

$$v = \gamma T \quad (37)$$

Define

$$G(z) = e^{-vh\left(\frac{z}{vh_1}\right)} \quad (38)$$

Because $h(z)$ is a Bernstein function, it follows that $G(z)$ is completely monotonic on the interval $0 \leq z < \infty$. Also we find

$$G(0) = e^{-vh(0)} = 1 \quad (39)$$

$$-G^{(1)}(0) = \frac{v}{vh_1} h^{(1)}\left(\frac{z}{vh_1}\right) e^{-vh\left(\frac{z}{vh_1}\right)}\Big|_{z=0} = 1 \quad (40)$$

Hence $G(z)$ may be written as in (13) where $F_\tau(t)$ is the CDF of a non-negative random variable $\tau$ with $E[\tau] = 1$. In this way we see that the normalized discrete process $\tau_{\overline{N_T}}(t)$ converges in distribution to a process $\tau_T(t)$ with univariate statistics described by the CDF $F_\tau$.

The path behavior of $\tau_T(t)$ depends on whether $\lim_{\kappa \to \infty} h(\kappa)$ is finite or infinite. Consider first the case where the limit is finite:

$$\lim_{\kappa \to \infty} h(\kappa) = C < \infty \quad (41)$$

The intensity of the underlying Poisson process becomes

$$\lim_{\kappa \to \infty} \lambda(\kappa) = \gamma C \quad (42)$$

In $\phi_K(u; \kappa)$ let $u = e^{-\frac{z}{K}}, 0 \leq z < \infty$:

$$\phi_K\left(e^{-\frac{z}{K}}; \kappa\right) = E\left[e^{-\frac{K}{K}z}\right] = 1 - \frac{h\left(\kappa\left(1-e^{-\frac{z}{K}}\right)\right)}{h(\kappa)}, 0 \leq z < \infty \quad (43)$$

Examine

$$\lim_{\overline{K} \to \infty} E\left[e^{-\frac{K}{K}z}\right] = 1 - \lim_{\kappa \to \infty} \frac{h\left(\kappa\left(1-e^{-\frac{h(\kappa)}{\kappa} \frac{z}{h_1}}\right)\right)}{h(\kappa)} = 1 - \frac{h\left(\frac{C}{h_1}z\right)}{C} \quad (44)$$

Define

$$g(z) = 1 - \frac{h\left(\frac{C}{h_1}z\right)}{C} \quad (45)$$

We find

$$g(0) = 1 \tag{46}$$

$$(-1)^n g^{(n)}(z) = (-1)^{n+1} \frac{C^{n-1}}{h_1^n} h^{(n)}\left(\frac{C}{h_1} z\right) \geq 0 \tag{47}$$

It follows that $g(z)$ is completely monotonic on $z \geq 0$ and may be written as

$$g(z) = \int_0^\infty e^{-zs} dF_\xi(s) \tag{48}$$

where $F_\xi(s)$ is the CDF of a non-negative random variable $\xi$. Thus $\frac{K}{\overline{K}}$ converges in distribution to $\xi$ as $\overline{K} \to \infty$:

$$\lim_{\overline{K} \to \infty} E\left[e^{-\frac{K}{\overline{K}} z}\right] = \int_0^\infty e^{-zs} dF_\xi(s) \tag{49}$$

The mean value of $\xi$ is given by

$$-g^{(1)}(0) = \int_0^\infty s \, dF_\xi(s) = (-1)^2 \frac{C^0}{h_1} h^{(1)}(0) = 1 \tag{50}$$

In this case $\tau_T(t)$ is a windowed compound Poisson process where the underlying Poisson process has intensity $\gamma C$ and the compounding variable $\xi$ has CDF $F_\xi(s)$.

Now consider the case where the limit is infinite:

$$\lim_{\kappa \to \infty} h(\kappa) = \infty \tag{51}$$

In this case the intensity of the underlying Poisson process also becomes infinite:

$$\lim_{\kappa \to \infty} \lambda(\kappa) = \infty \tag{52}$$

Following the same analysis as above and using (21) we find

$$\lim_{\overline{K} \to \infty} E\left[e^{-\frac{K}{\overline{K}} z}\right] = 1 - \lim_{\kappa \to \infty} \frac{h\left(h(\kappa)\frac{z}{h_1}\right)}{h(\kappa)} = 1 \tag{53}$$

Hence we may write

$$\lim_{\overline{K} \to \infty} E\left[e^{-\frac{K}{\overline{K}} z}\right] = \int_0^\infty e^{-zs} dF_\xi(s) \tag{54}$$

where

$$F_\xi(s) = U(s) \tag{55}$$

is the unit step function. In this case $\frac{K}{\overline{K}}$ converges in distribution to a random variable $\xi$ with all of its mass at $\xi = 0$. When $\lim_{\kappa \to \infty} h(\kappa) = \infty$, the process $\tau(t)$ is an infinite activity process and cannot be realized as a windowed compound Poisson process. Instead the windowed compound Poisson process $\tau_{\overline{N_T}}(t)$ only approximates the desired process $\tau(t)$.

In summary, the procedure is as follows:

1. Choose $\gamma > 0, T > 0$. Define $\nu = \gamma T$.

2. Choose a Bernstein function $h(z)$ satisfying the conditions above. It can either be chosen directly or it can be chosen as $h(z) = -\frac{1}{\nu}\ln G(\nu h_1 z)$ where $G(z)$ is the PGF of an infinitely divisible non-negative random variable $\tau$ having unit mean and finite variance. In this latter case it is convenient to choose $h_1 = 1$.

3. If $\lim_{z \to \infty} h(z) = C$:

    a. Define $g(z) = 1 - \frac{h\left(\frac{C}{h_1}z\right)}{C}$. Generate positive random variables $\xi$ having PGF $g(z)$.

    b. Generate a Poisson process with intensity $\lambda = \gamma C$.

    c. Generate a windowed compound-Poisson process of length $T$ from the Poisson process and the normalized random variables $\frac{\xi}{\nu}$ where $\nu = \lambda T = \gamma C T$. This process is the desired finite activity process $\tau_T(t)$.

4. If $\lim_{z \to \infty} h(z) = \infty$:

    a. Choose $\kappa \gg 0$ and generate discrete random integers with PMF $p_K(n; \kappa)$.

    b. Generate a Poisson process with intensity $\lambda = \gamma h(\kappa)$.

    c. Generate a windowed integer-valued compound-Poisson process from the Poisson process and the discrete random variables $K_i, i = 1, 2, \cdots$.

    d. Normalize the windowed compound-Poisson process by the value $\lambda T \overline{K}$. This normalized process approximates the desired infinite activity process $\tau(t)$.

**Covariance of $\tau(t)$**

Consider the covariance of $\tau(t)$ given by

$$Cov_\tau(s) = E[\tau(t+s)\tau(t)] - 1 \qquad (56)$$

To determine $Cov_\tau(s)$ it is helpful to examine Figure 2:

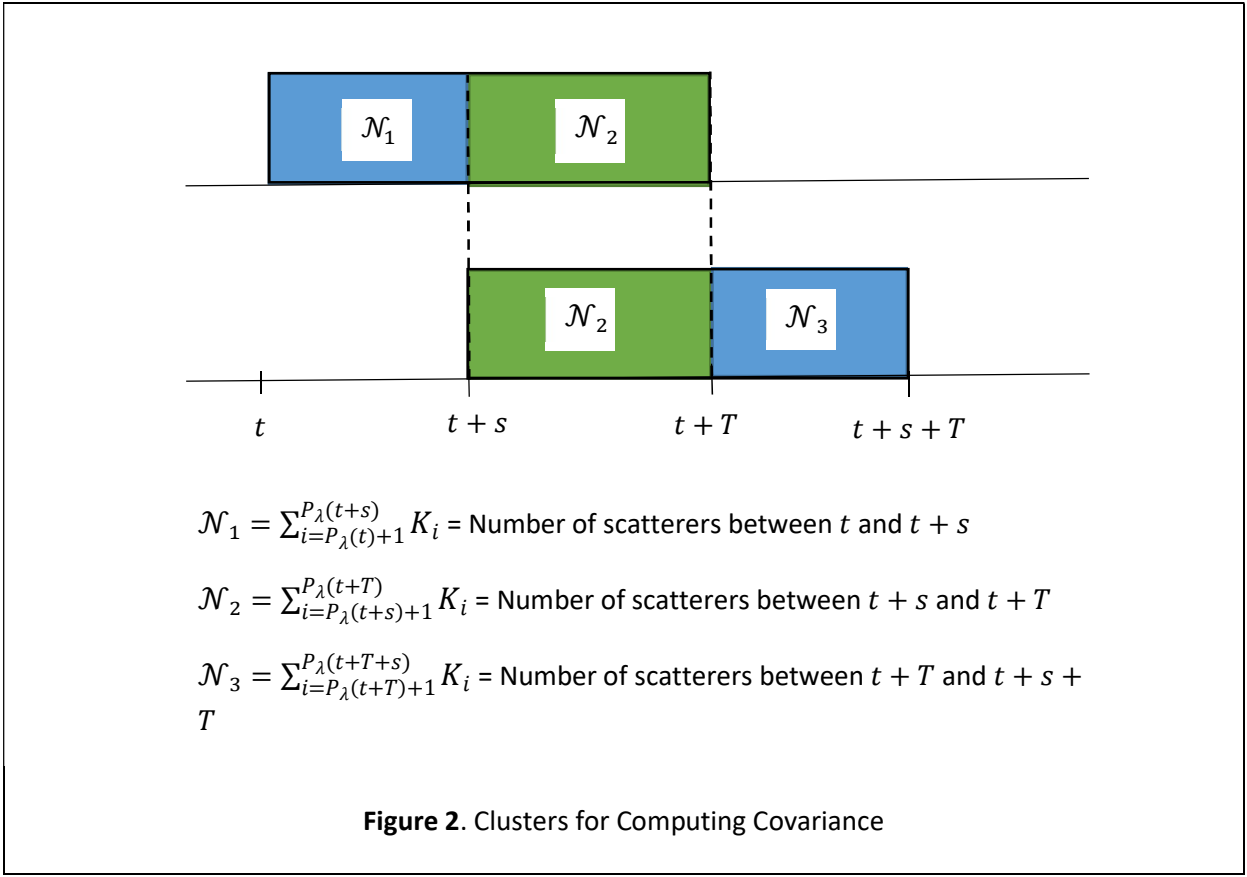

$\mathcal{N}_1 = \sum_{i=P_\lambda(t)+1}^{P_\lambda(t+s)} K_i$ = Number of scatterers between $t$ and $t+s$

$\mathcal{N}_2 = \sum_{i=P_\lambda(t+s)+1}^{P_\lambda(t+T)} K_i$ = Number of scatterers between $t+s$ and $t+T$

$\mathcal{N}_3 = \sum_{i=P_\lambda(t+T)+1}^{P_\lambda(t+T+s)} K_i$ = Number of scatterers between $t+T$ and $t+s+T$

**Figure 2**. Clusters for Computing Covariance

This figure shows two different positions of the window of length $T$ at times $t$ and $t+s$. We have labeled the numbers of scatterers in various parts of the window as $\mathcal{N}_1, \mathcal{N}_2$, and $\mathcal{N}_3$. Thus the number of scatterers in the window at times $t$ and $t+s$ are given by

$$N_T(t) = \mathcal{N}_1 + \mathcal{N}_2 \qquad (57)$$

$$N_T(t+s) = \mathcal{N}_2 + \mathcal{N}_3 \qquad (58)$$

We may now write

$$E[N_T(t+s)N_T(t)] = E[\mathcal{N}_1(\mathcal{N}_2 + \mathcal{N}_3)] + E[\mathcal{N}_2(\mathcal{N}_2 + \mathcal{N}_3)] \qquad (59)$$

Because a Levy process has independent increments, the window between $t$ and $t + s$ is independent of the window between $t + s$ and $t + T$. Thus the first term on the right-hand side becomes

$$E[\mathcal{N}_1(\mathcal{N}_2 + \mathcal{N}_3)] = E[\mathcal{N}_1]E[(\mathcal{N}_2 + \mathcal{N}_3)] = (\lambda s \overline{K})(\lambda T \overline{K}) \tag{60}$$

The second term on the right-hand side may be written as

$$E[\mathcal{N}_2(\mathcal{N}_2 + \mathcal{N}_3)] = E[(\mathcal{N}_2)^2] + E[\mathcal{N}_2 \mathcal{N}_3] \tag{61}$$

The window between $t + s$ and $t + T$ is independent of the window between $t + T$ and $t + s + T$. Hence we have

$$E[\mathcal{N}_2 \mathcal{N}_3] = E[\mathcal{N}_2]E[\mathcal{N}_3] = (\lambda(T-s)\overline{K})(\lambda s \overline{K}) \tag{62}$$

For the remaining term we find

$$E[(\mathcal{N}_2)^2] = E\left[\left(\sum_{i=P_\lambda(t+s)+1}^{P_\lambda(t+T)} K_i\right)^2\right] = E[P_\lambda(t+T)(P_\lambda(t+T) - 1)]\overline{K}^2 + E[P_\lambda(t+T)]E[K^2] \tag{63}$$

From the probability generating function of $K$ we find

$$E[K^2] = -\frac{\kappa^2}{h(\kappa)}h_2 + \overline{K} = \frac{-\kappa^2 h_2 + \kappa}{h(\kappa)} \tag{64}$$

Also we have

$$E[P_\lambda(t+T)(P_\lambda(t+T) - 1)] = (\lambda(T-s))^2 \tag{65}$$

$$E[P_\lambda(t+T)] = \lambda(T-s) \tag{66}$$

Combining these results now yields

$$E[N_T(t+s)N_T(t)] = \overline{N_T}^2 + \overline{N_T}\left(1 - \frac{s}{T}\right)(1 - \kappa h_2) \tag{67}$$

We now may write

$$E\left[\frac{N_T(t+s)}{\overline{N_T}} \frac{N_T(t)}{\overline{N_T}}\right] - 1 = \left(\frac{-h_2}{\nu} + \frac{1}{\overline{N_T}}\right)\left(1 - \frac{s}{T}\right) \tag{68}$$

This result assumes $s \leq T$. If $s > T$ then $N_T(t)$ and $N_T(t + s)$ are independent and we find

$$E\left[\frac{N_T(t+s)}{\overline{N_T}} \frac{N_T(t)}{\overline{N_T}}\right] - 1 = 0 \tag{69}$$

In the limit as $\overline{N_T} \to \infty$ we now find:

$$Cov_\tau(s) = \begin{cases} \frac{-h_2}{\nu}\left(1 - \frac{s}{T}\right), & 0 \leq s \leq T \\ 0, & s > T \end{cases} \tag{70}$$

For $s > T$ the values $\tau(t + s)$ and $\tau(t)$ are independent as they essentially represent two distinct increments of a compound-Poisson process. The variance of $\tau(t)$ is

$$Var_\tau = \frac{-h_2}{\nu} \tag{71}$$

As $v$ decreases the fluctuations in $\tau$ increase. Therefore the number $v$ plays the role of a shape parameter whose value determines the degree of non-Gaussianity of the scattered signal. Because $v$ and $T$ may be varied independently through choice of $\gamma$, it is possible to have a process $\tau(t)$ with both high variance and long correlation by choosing $v$ and $T$ appropriately.

As illustrated in Figure 1, physically one may think of an integer-valued compound-Poisson process as specifying clusters where the clusters occur in accordance with a Poisson process of intensity $\lambda$ and the number of scatterers in each cluster is a random number $K$. The average number of clusters in a unit time period is therefore equal to $\lambda$. Define

$$T_{cluster} = \frac{1}{\lambda} = \frac{1}{\gamma h(\kappa)} \tag{72}$$

Heuristically one may think of $T_{cluster}$ as the average period of time between clusters. In the finite activity case we have in the limit

$$v = \gamma T = \frac{\lambda}{C} T = \frac{1}{C} \frac{T}{T_{cluster}} \tag{73}$$

This observation suggests the following physical interpretation: $v$ measures the size of the window $T$ in units proportional to the average time between clusters $T_{cluster}$. We expect that as the window $T$ grows large relative to the average time between clusters, i.e. as the window encompasses an increasingly large number of clusters, the statistics of the scattered field will tend toward Gaussian. To see this result explicitly let $v \to \infty$ in (38) and use (22, 24-25):

$$\lim_{v\to\infty} G(z) = \lim_{v\to\infty} e^{-vh\left(\frac{z}{vh_1}\right)} = e^{-z} \tag{74}$$

We thus see that as $v \to \infty$ the distribution of random variable $\tau$ tends to

$$F_\tau(\tau) = U(\tau - 1) \tag{75}$$

where $U(\tau)$ is the unit step function. The compound-Gaussian model in (10) reduces to a Gaussian model in this case.

**Example: A Finite Activity Process**

Let $h(z)$ be

$$h(z) = \frac{z}{z+1} \tag{76}$$

We have

$$(-1)^n h^{(n+1)}(z) = (n+1)!\left(\frac{1}{z+1}\right)^{n+2} \geq 0, n = 0, 1, 2, \cdots \tag{77}$$

Therefore $h^{(1)}(z)$ is completely monotonic and $h(z)$ is a Bernstein function. It is straightforward to verify that $h(z)$ in (76) satisfies (21-22, 24-25) with $h_1 = 1$. The PMF $p_K(n; \kappa)$ becomes

$$p_K(n;\kappa) = \begin{cases} p(1-p)^{n-1}, & n = 1,2,\cdots \\ 0, & n = 0 \end{cases} \tag{78}$$

where we have defined

$$p = \frac{1}{\kappa+1} \tag{79}$$

This is the PMF of a geometrically distributed random variable. The probability generating function is given by

$$\phi_K(u;\kappa) = 1 - \frac{h(\kappa(1-u))}{h(\kappa)} = \frac{pu}{1-(1-p)u} \tag{80}$$

The probability generating function of $N_T$ becomes

$$\phi_{N_T(t)}(u) = e^{-\nu\left(\frac{1-u}{1-(1-p)u}\right)} \tag{81}$$

In this case $N_T$ is known to have a Polya-Aeppli distribution with PMF [21]:

$$p_{N_T}(n) = \begin{cases} \sum_{k=1}^{n} e^{-\gamma(1-p)} \frac{(\gamma(1-p))^k}{k!} \binom{n-1}{k-1}(1-p)^{n-k}p^k, & n = 1,2,\cdots \\ e^{-\nu h(\kappa)}, & n = 0 \end{cases} \tag{82}$$

The limiting function $G(z)$ becomes

$$G(z) = e^{-\nu\frac{z}{z+\nu}} \tag{83}$$

The intensity of the underlying Poisson process is given by

$$\lambda = \gamma(1-p) \tag{84}$$

As $\kappa \to \infty$ we have $p \to 0$ and hence the intensity $\lambda \to \gamma$ remains finite. We may therefore write

$$g(z) = 1 - h(z) = \frac{1}{z+1} = \int_0^\infty e^{-z\xi} dF_\xi(\xi) \tag{85}$$

with

$$f_\xi(\xi) = \frac{dF_\xi(\xi)}{d\xi} = e^{-\xi}, \xi \geq 0 \tag{86}$$

The limiting process $\tau(t)$ may thus be realized exactly as a compound-Poisson process with an exponentially distributed mixing variable $y = \frac{\xi}{\nu}$ with PDF

$$f_y(y) = \nu e^{-\nu y}, y \geq 0 \tag{87}$$

In particular, one may generate a Poisson process with intensity $\gamma$ and use this process as the basis of the process in (30) with exponential mixing variables having mean $\frac{1}{\nu}$ where $\nu$ is given by (73). The PDF of $\tau$ can be shown to be

$$f_\tau(\tau) = e^{-\nu}\delta(\tau) + (1 - e^{-\nu})g_\tau(\tau) \tag{88}$$

where

$$g_\tau(\tau) = \frac{\nu e^{-\nu}}{1-e^{-\nu}}\frac{e^{-\nu\tau}}{\sqrt{\tau}}I_1(2\nu\sqrt{\tau}) \tag{89}$$

In this expression $\delta(\tau)$ denotes the Dirac delta function and $I_1(z)$ denotes the modified Bessel function. The non-zero mass at $\tau = 0$ occurs because there is a non-zero probability $e^{-\nu}$ that the window will not encompass any scatterers (i.e. will be wholly between scatterers). The PDF $g_\tau(\tau)$ is therefore the PDF of $\tau$ conditioned on at least one scatterer being in the window.

Figures 3-7 show plots of sample paths of the limiting process $\tau(t)$ for this example. The units of time are arbitrary. For these plots $\gamma = 0.25$ and $\nu$ was varied by changing the length $T$ of the window. Also shown are histograms as well as a comparison of the histograms to $(1 - e^{-\nu})g_\tau(\tau)$ in (88-89) (we do not plot the Dirac delta contribution to the predicted PDF although this contribution is included in the histograms – it is especially clear in Figures 6 and 7).

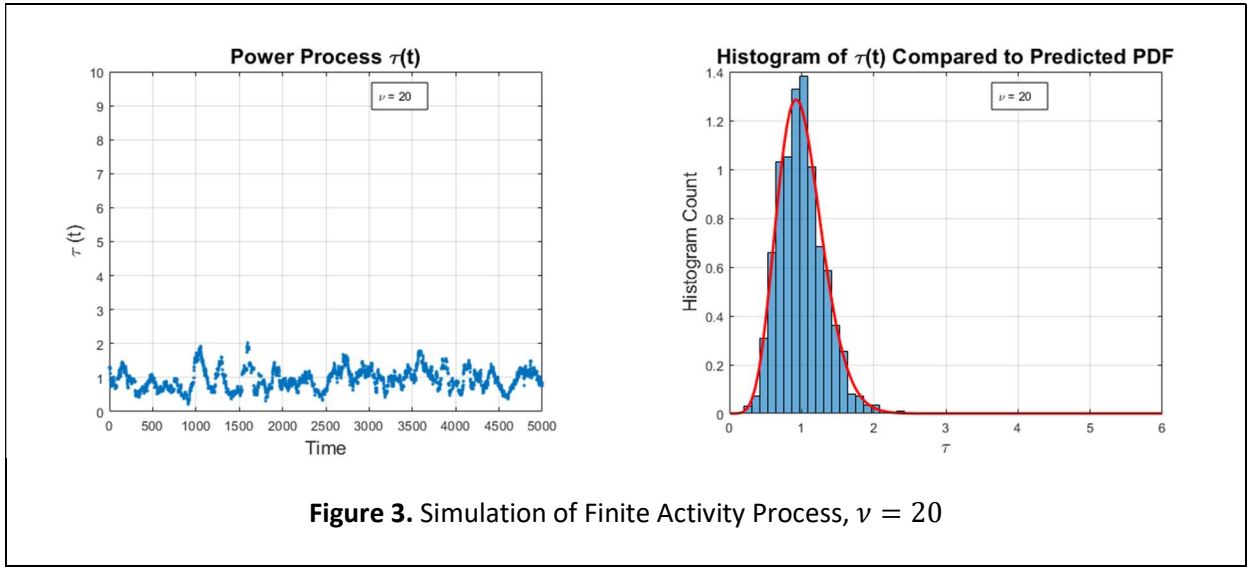

**Figure 3.** Simulation of Finite Activity Process, $\nu = 20$

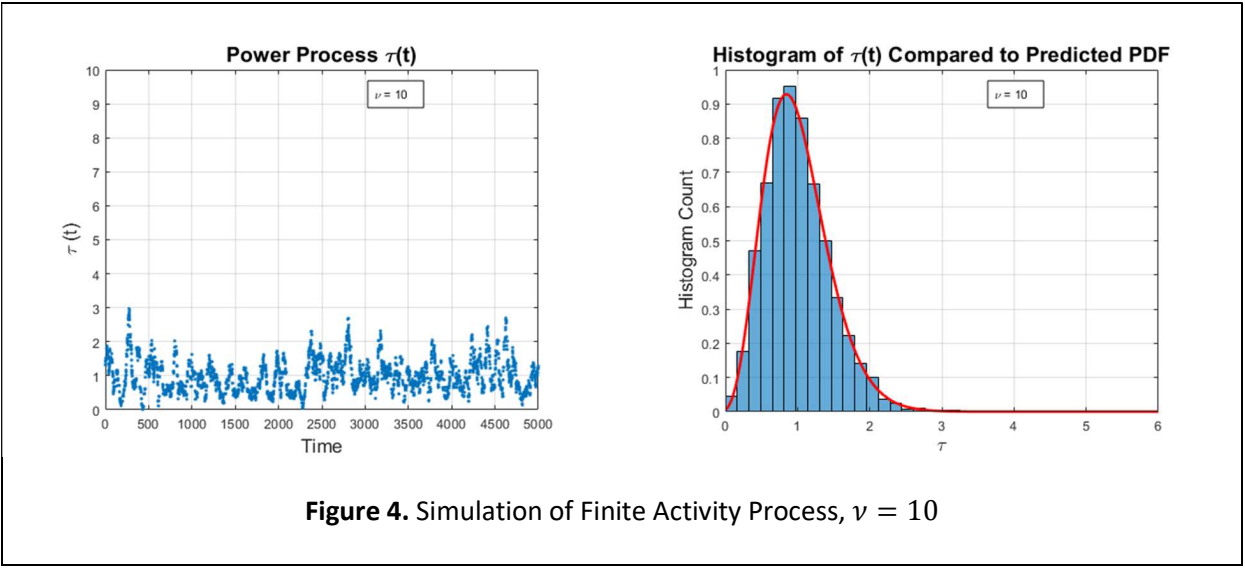

**Figure 4.** Simulation of Finite Activity Process, $\nu = 10$

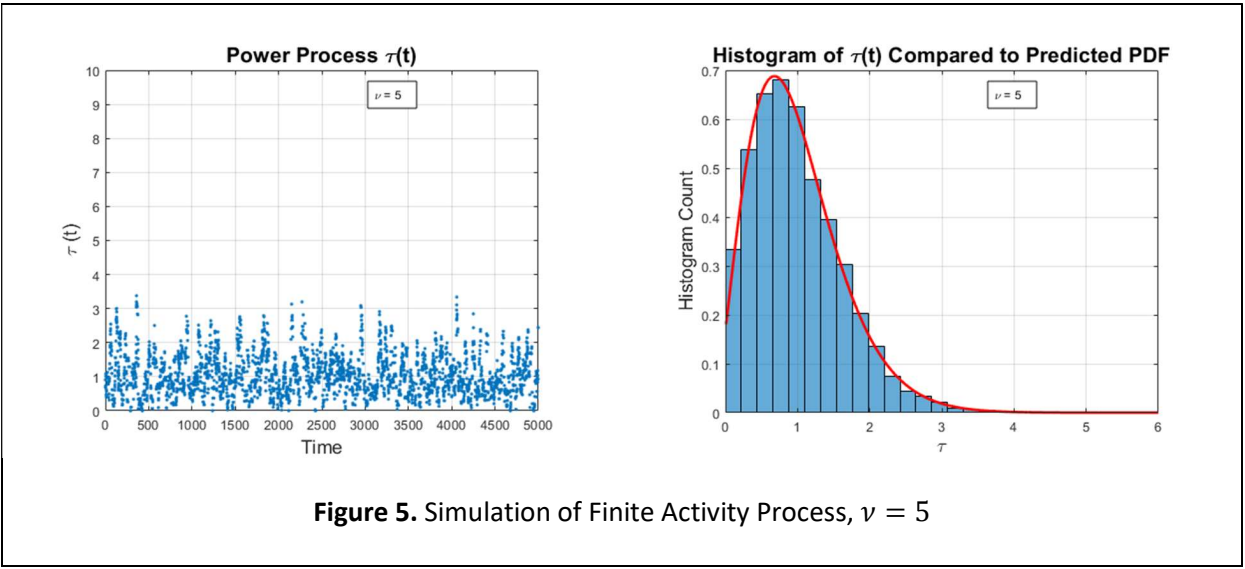

**Figure 5.** Simulation of Finite Activity Process, $\nu = 5$

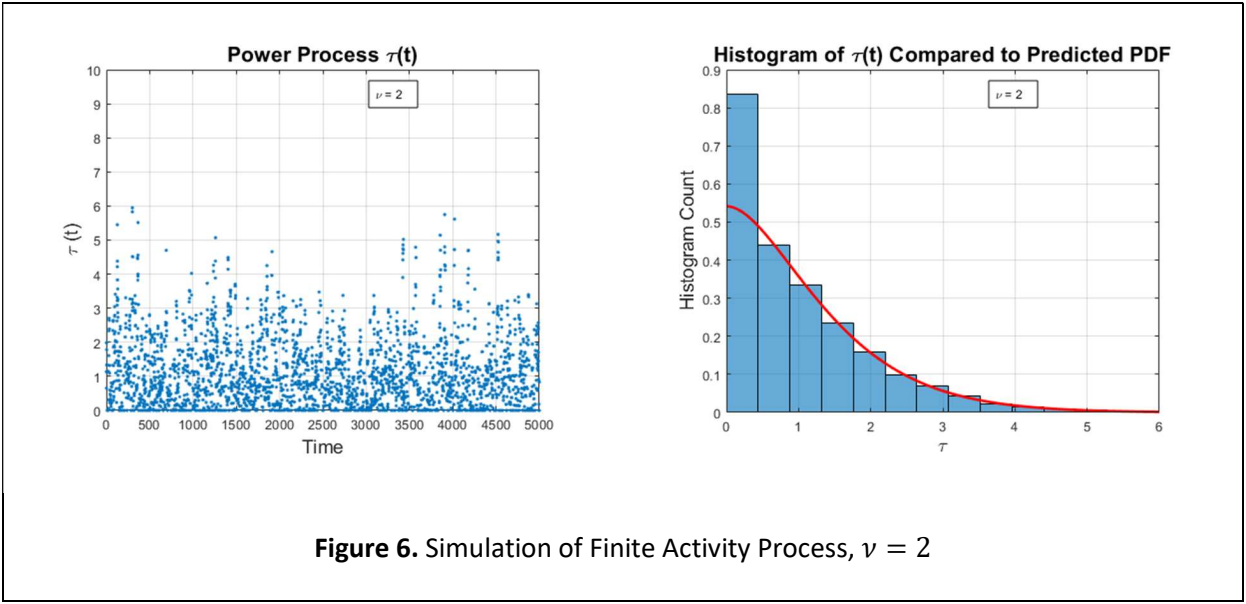

**Figure 6.** Simulation of Finite Activity Process, $\nu = 2$

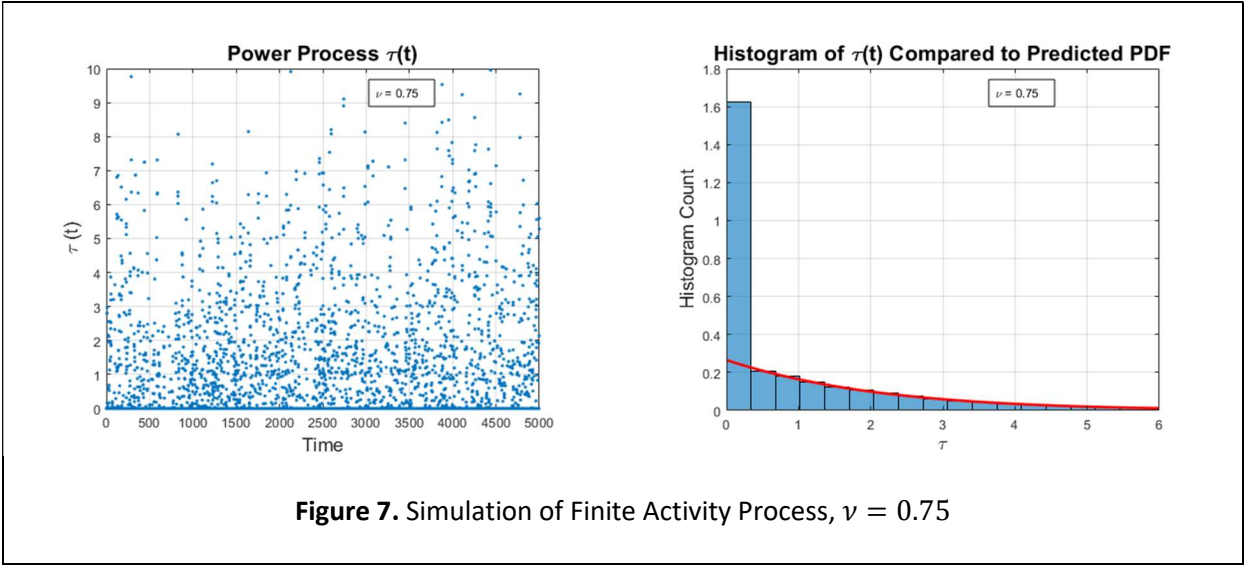

**Figure 7.** Simulation of Finite Activity Process, $\nu = 0.75$

In this example we have $-h_2 = 2$. Thus the variance is

$$Var_\tau = \frac{2}{v} \qquad (90)$$

As is clearly shown in these figures, as $v$ decreases the fluctuations in the power process $\tau(t)$ increase. Increasing the fluctuations in $\tau(t)$ in turn induces increased non-Gaussianity in the scattered field.

Figure 8 shows the results illustrated in Figure 7 over both small and large time scales.

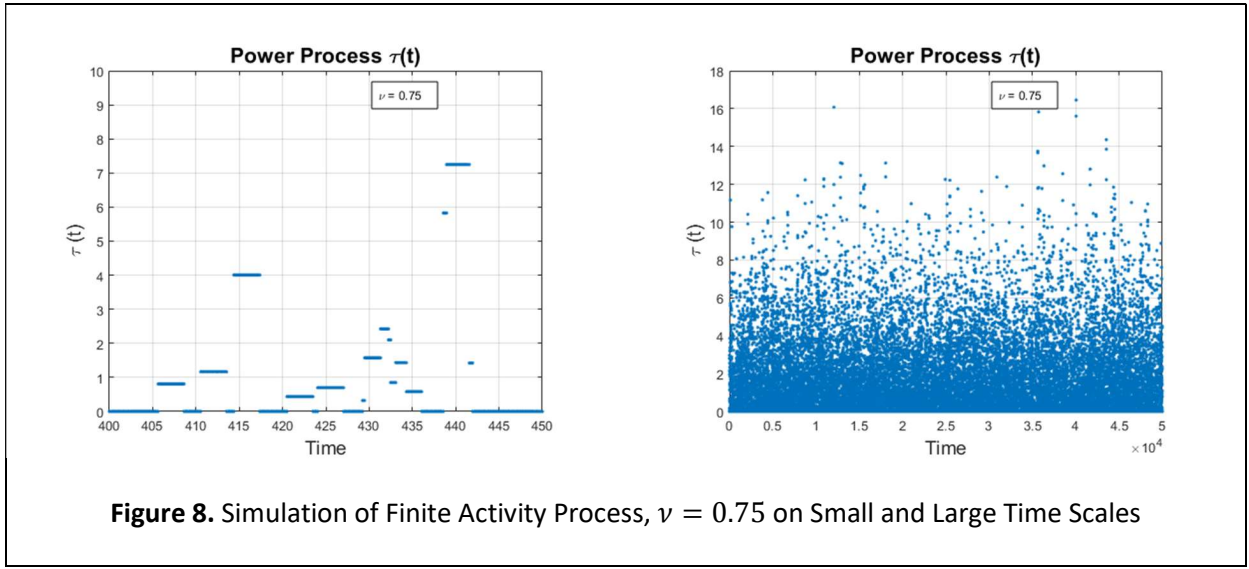

**Figure 8.** Simulation of Finite Activity Process, $v = 0.75$ on Small and Large Time Scales

Figure 8 shows that on a small time scale the process is constant for random periods of time. This occurs because the number of scatterers within the window does not change until either a cluster leaves the window or a new cluster enters the window. Until one of these events occurs, the number of scatterers within the window is constant and hence $\tau(t)$ remains constant. Also, on the small time scale Figure 7 shows that for small values of $v$, the process is often zero. This occurs if the window is between clusters, which occurs increasingly often as the window size gets small compared to the average time between clusters. These zero values give rise to the mass at zero shown in the PDF in (88) and reflected in the histograms in Figures 7 and 8. Over a large time scale, Figure 9 shows the possibility of relatively large values of the power $\tau$. This behavior would lead to extreme non-Gaussianity in the resulting compound-Gaussian process.

**Example: An Infinite Activity Process**

Let $h(z)$ be

$$h(z) = \ln(1+z) \tag{91}$$

We have

$$(-1)^n h^{(n+1)}(z) = n!\left(\frac{1}{1+z}\right)^{n+1} \geq 0, n = 0,1,2,\cdots \tag{92}$$

Therefore $h^{(1)}(z)$ is completely monotonic and $h(z)$ is a Bernstein function. It is straightforward to show that $h(z)$ in (91) satisfies (21-22, 24-25) with $h_1 = 1$. The PMF $p_K(n;\kappa)$ becomes

$$p_K(n;\kappa) = \begin{cases} -\frac{1}{\ln(1-p)}\frac{p^n}{n}, & n = 1,2,\cdots \\ 0, & n = 0 \end{cases} \tag{93}$$

where we have defined

$$p = \frac{\kappa}{1+\kappa} = \frac{\overline{N_T}}{\nu + \overline{N_T}} \tag{94}$$

This is the PMF of a logarithmically distributed random variable. The PMF $p_{N_T}(n;\overline{N_T})$ becomes

$$p_{N_T}(n;\overline{N_T}) = \frac{\Gamma(n+\nu)}{\Gamma(\nu)\Gamma(n+1)} \frac{\left(\frac{\overline{N_T}}{\nu}\right)^n}{\left(1+\frac{\overline{N_T}}{\nu}\right)^{\nu+n}}, n = 0,1,2,\cdots \tag{95}$$

This is the PMF of a negative binomial distributed random variable. It is known that the compound-Poisson random variable formed with a logarithmically distributed random variable has a negative binomial distribution.

The limiting function $G(z)$ is now given by

$$G(z) = e^{-\nu \ln\left(1+\frac{z}{\nu}\right)} = \left(\frac{1}{1+\frac{z}{\nu}}\right)^\nu = \int_0^\infty e^{-z\tau} dF_\tau(\tau) \tag{96}$$

where

$$f_\tau(\tau) = \frac{dF_\tau(\tau)}{d\tau} = \frac{\nu^\nu}{\Gamma(\nu)}\tau^{\nu-1}e^{-\nu\tau} \tag{97}$$

is the probability density function (PDF) of a gamma distributed random variable $\tau$. Thus the normalized compound-Poisson process formed from logarithmically distributed random integers approximates a gamma distributed process when $\overline{N_T} \gg 1$. The intensity of the underlying Poisson process $\lambda = \gamma \ln\left(1 + \frac{\overline{N_T}}{\nu}\right)$ goes to infinity as $\overline{N_T} \to \infty$ and is consistent with the limit process $\tau(t)$ being an infinite activity process, which the gamma Levy process is known to be [22].

Figures 9-13 show plots of sample paths of the process $\tau_{\overline{N_T}}(t)$ for this example. For these plots $\gamma = 0.25$, $\kappa = 150$, and the value of $\nu$ was varied by changing the length $T$ of the window. Also shown are the corresponding sample PDF and CDF of $\tau_{\overline{N_T}}$ compared to the gamma distribution (sample results are the blue lines and the gamma PDF and CDF are plotted in red):

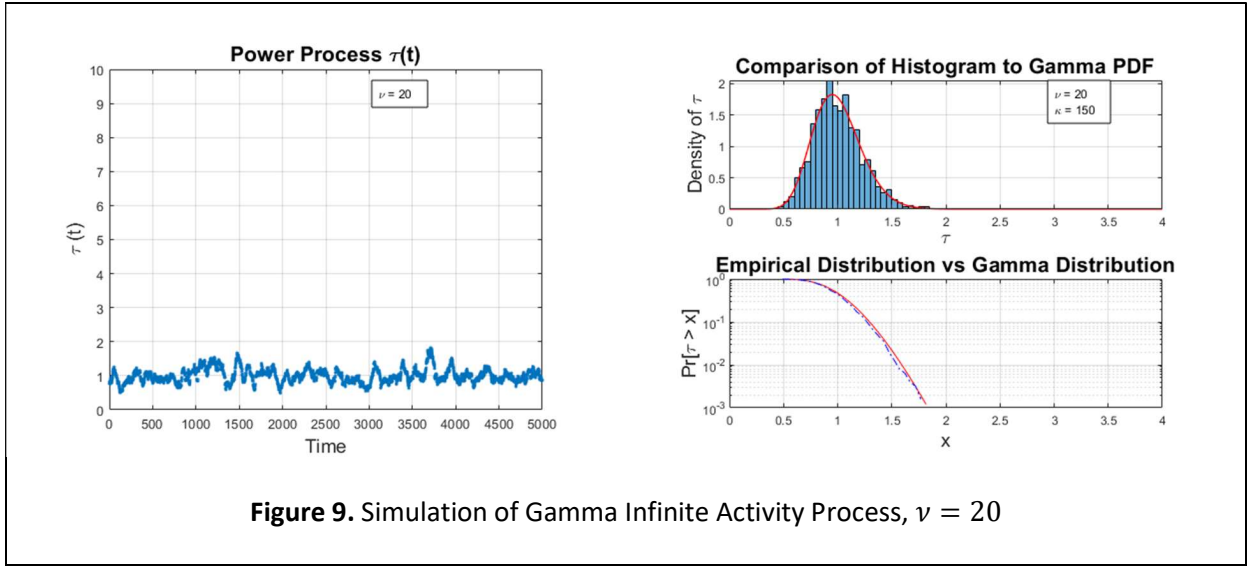

**Figure 9.** Simulation of Gamma Infinite Activity Process, $\nu = 20$

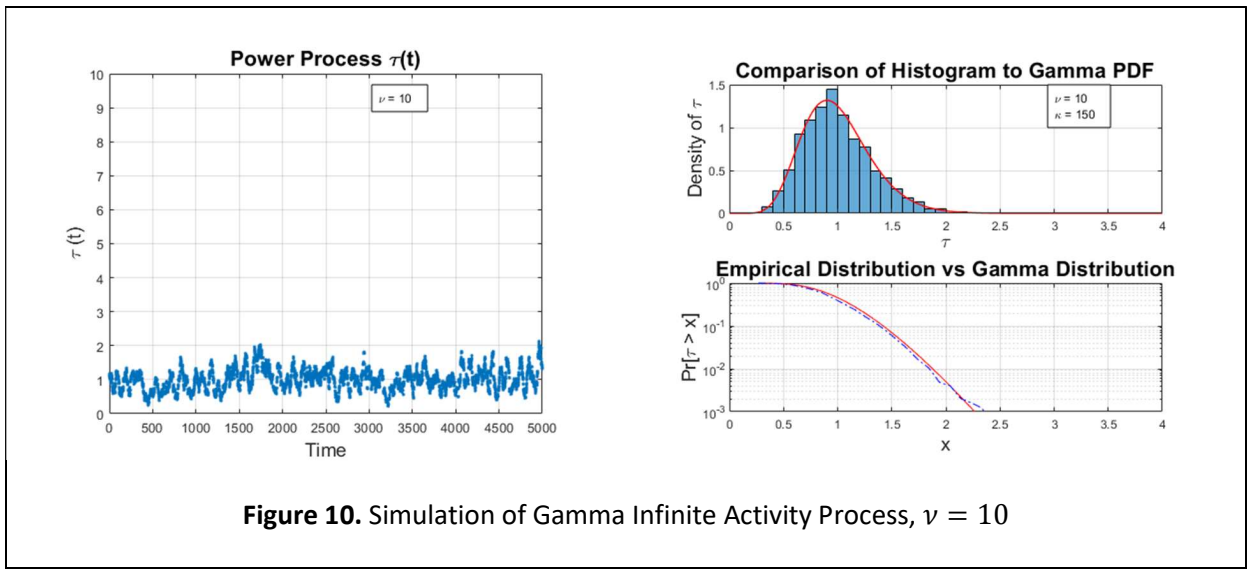

**Figure 10.** Simulation of Gamma Infinite Activity Process, $\nu = 10$

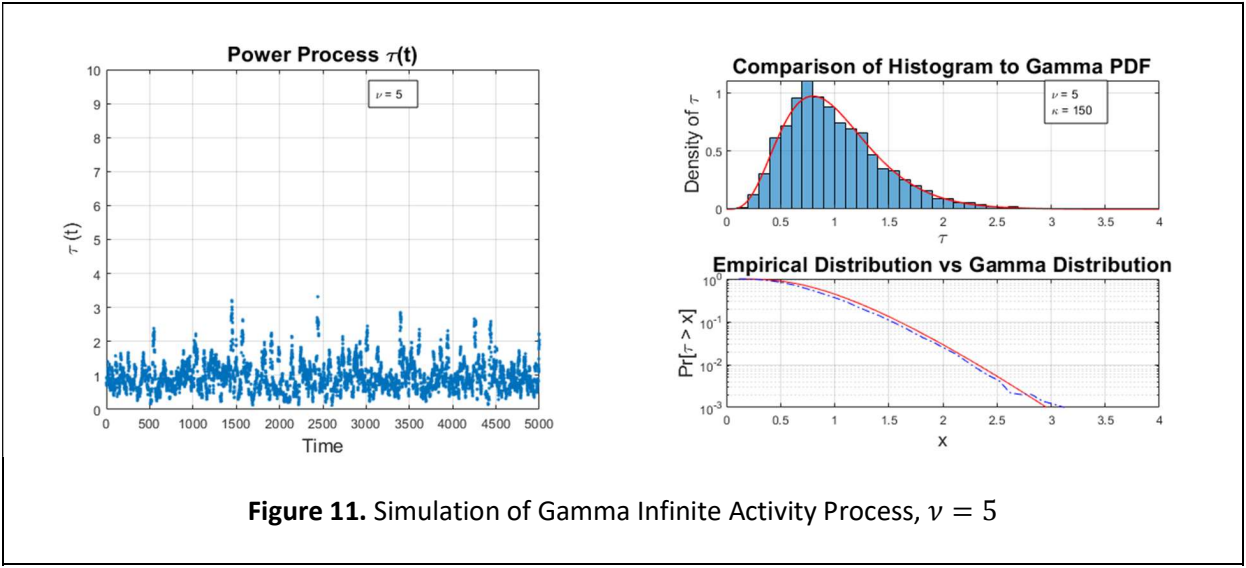

**Figure 11.** Simulation of Gamma Infinite Activity Process, $\nu = 5$

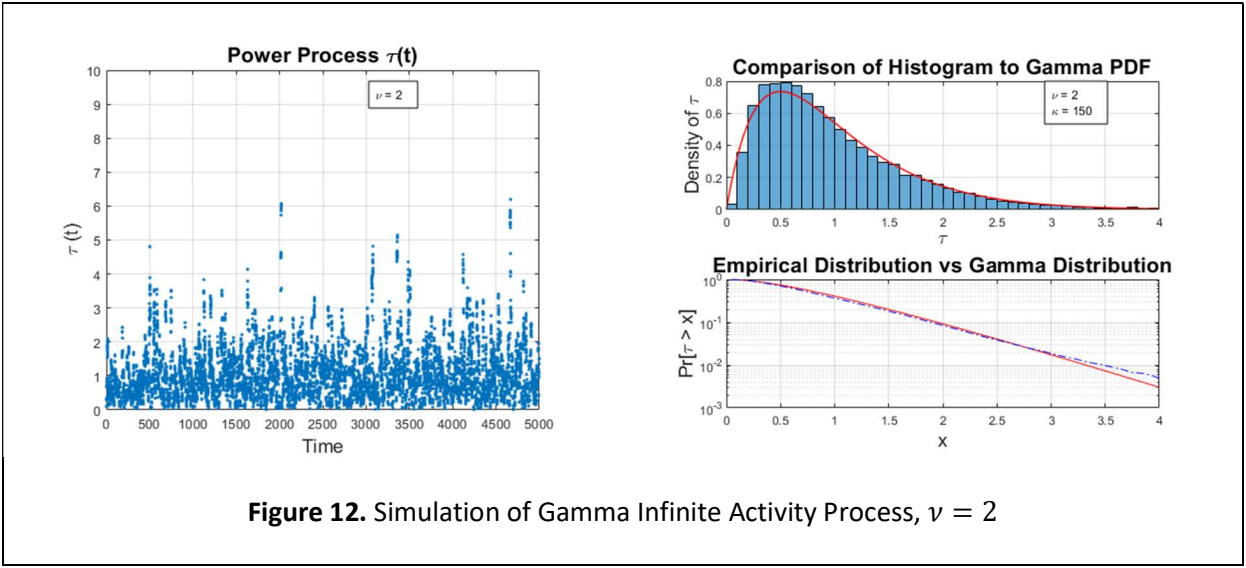

**Figure 12.** Simulation of Gamma Infinite Activity Process, $\nu = 2$

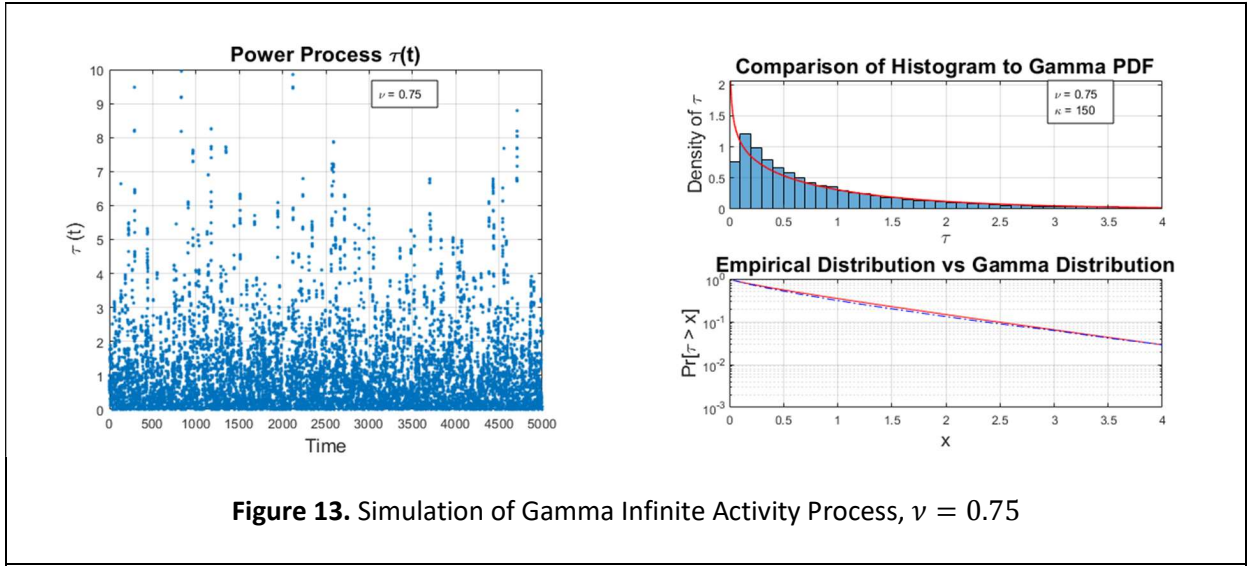

**Figure 13.** Simulation of Gamma Infinite Activity Process, $\nu = 0.75$

In this example we have $-h_2 = 1$ and thus the variance is

$$Var_\tau = \frac{1}{\nu} \tag{98}$$

This is the variance of a gamma random variable with PDF in (97). As with the finite activity process, as $\nu$ decreases the fluctuations of $\tau(t)$ increase and the non-Gaussianity of the scattered field increases.

Figure 14 shows the example in Figure 13 over small and large time scales.

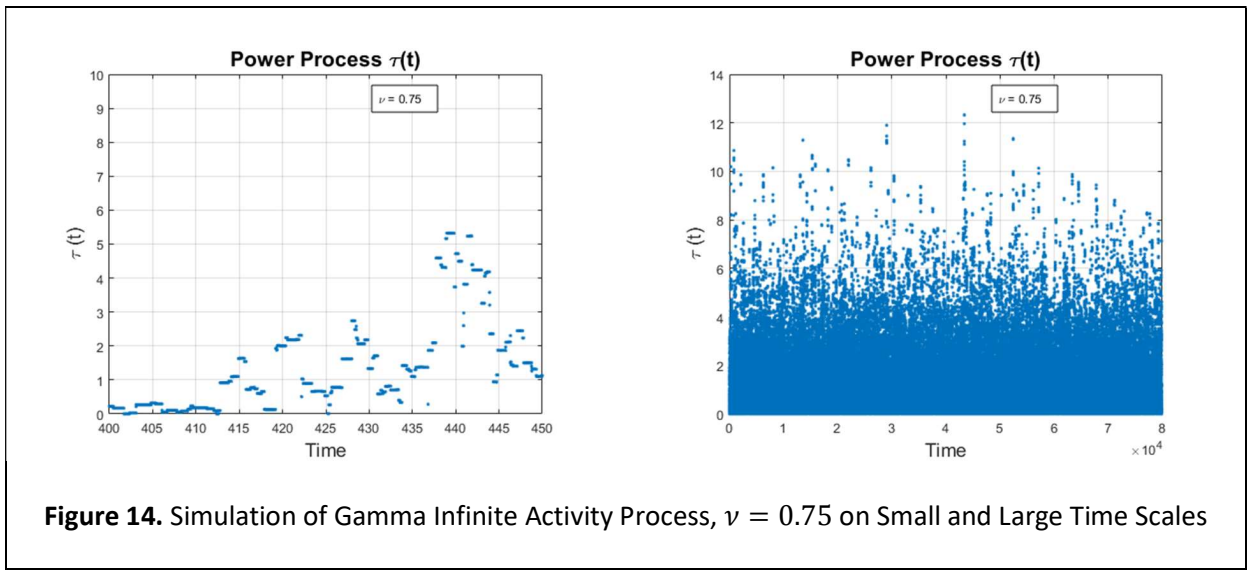

**Figure 14.** Simulation of Gamma Infinite Activity Process, $\nu = 0.75$ on Small and Large Time Scales

On a small time scale this process does not equal zero (in the limit). This behavior differs signficantly from the finite activity process shown in Figure 9. On the other hand, over a large time scale this example can have large values and will lead to extreme deviation from Gaussianity in the corresponding compound-Gaussian process.

## Conclusion

In this work we used the theory of Bernstein functions, completely monotonic functions, and Levy processes to develop a random process model for a non-negative infinitely divisible process $\tau(t)$. In particular, we used a Bernstein function satisfying certain conditions and having a specified relationship to the Laplace-Stieltjes transform of the distribution $F_\tau(\tau)$ of an infinitely divisible random variable $\tau > 0$ to define a discrete random variable. We then used this discrete random variable as a mixing variable to define a windowed compound-Poisson process. The normalized version of this windowed process then leads to a model of $\tau(t)$ having univariate statistics described by $F_\tau(\tau)$.

Several avenues of further development remain to be explored. The examples herein used the fact that the derivatives $h^{(n)}(z)$ were available in closed-form to describe the PMF of the mixing random variable $K$. Although a large number of functions are known to be Bernstein functions [15] and hence potentially can lead to a large number of compound-Gaussian process models, it is not always possible to obtain their derivatives in closed-form or in a manner that provides for ease of computation. However, as shown in equation (26) the probability generating function of the mixing random variable is available directly from $h(z)$. Therefore, methods for generating random variates directly from a probability generating function should be helpful for simulating a wide class of compound-Gaussian processes [23-24]. In another direction of possible development, [25-27] give a very interesting approach to modeling the scattered field in terms of stochastic differential equations. Because an extensive stochastic calculus is available with respect to Levy processes [22, 28], it would be interesting to incorporate the integer-valued process defined herein into the approach of [25-27].

## References

[1] Trunk, G.V., "Radar Properties of Non-Rayleigh Sea Clutter," *IEEE Trans. on Aerospace and Electronic Systems*, Vol. 8, No. 2, March 1972, 196-204.

[2] K. D. Ward, "Compound representation of high resolution sea clutter," *Electron. Lett.*, vol. 17, no. 16, pp. 561–563, Aug. 1981.

[3] Jakeman, E., Pusey, P." A Model for Non-Rayleigh Echo," *IEEE Transactions on Antennas and Propagation*, Vol. AP-24, No. 6, November 1976, 806-14.

[4] Robbins, H., "The Asymptotic Distribution for Sums of a Random Number of Random Variables," *Bulletin of the American Mathematical Society*, Vol. 54, 1948, 1151-61.

[5] Gnedenko, B., Fahim, H., "On a Transfer Theorem," *Doklady Akademii Nauk SSSR*, Vol. 187, 1969, 15-17.

[6] Szasz, D., Freyer, B., "A Problem of Summation Theory with Random Indices," *Litovskii Matematicheskii Sbornik*, Vol. 11, 1971, 181-187.

[7] Szasz, D., "On the Limiting Cases of Distributions for Sums of a Random Number of Independent, Identically Distributed Random Variables," *Theory of Probability and Its Applications*, Vol. 17, 1972, 401-15.

[8] Szasz, D., ""Stability and Law of Large Numbers for Sums of a Random Number of Random Variables," *Acta Scientarium Mathematicarum*, Vol. 33, 1972, 269-74.

[9] Schoenberg I.J. 1938. "Metric Spaces and Completely Monotonic Functions," *Annals of Mathematics* 39(4), 811-41.

[10] Farschian, M., Posner, F.L., "The Pareto Distribution for Low Grazing Angle and High Resolution X-Band Sea Clutter," *Proceedings of the 201 IEEE International Radar Conference*, May 2010.

[11] Xue, J., Xu, S., Liu, J., Shui, P., "Model for Non-Gaussian Sea Clutter Amplitudes Using Generalized Inverse Gaussian Texture," *IEEE Geoscience and Remote Sensing Letters*, Vol. 16, No. 6, June 2019, 892-896.

[12] Miller, K., Samko, S., "Completely Monotonic Functions," *Integral Transforms and Special Functions*, Vol. 12, No. 4, 2001, 389-402.

[13] Merkle, M., "Completely Monotone Functions – A Digest," arXiv:1211.0900v1 [math.CA] 2 Nov. 2012.

[14] Koumandos, S., "On Completely Monotonic and Related Functions," in Rassias, T., Pardalos, P., (eds.) MATHEMATICS WITHOUT BOUNDARIES, Springer, 285-321, 2014.

[15] Schilling, R., Song, R., Vondracek, Z., BERNSTEIN FUNCTIONS: THEORY AND APPLICATIONS, De Gruyter, 2010.

[16] Widder, D.V., THE LAPLACE TRANSFORM, Dover Publications, 1941.

[17] Bochner, S., "Completely Monotone Functions of the Laplace Operator for Torus and Sphere," *Duke Mathematicl Journal*, Vol. 3, No. 9, 1937, 488-503.

[18] Hansen, B., "Monotonicity Properties of Infinitely Divisible Distributions," Technische Universiteit Eindhoven. https://doi.org/10.6100/IR293678, 1988.

[19] Jakeman, E., "On the Statistics of K-Distributed Noise," *Journal of Physics A: Mathematical and General*, Vol. 13, 1980, 31-48.

[20] Sangston, K. J. and Gerlach, K.R. "Coherent Detection of Radar Targets in a Non-Gaussian Background," IEEE Transactions on AES, Vol. 30, No. 2, April 1994, 330-340.

[21] Chukova, S., Minkova, L., "Characterization of the Polya-Aeppli Process," *Stochastic Analysis and Applications*, Vol. 34, No. 1, 2013, 590-99.

[22] Kyprianou, A., INTRODUCTORY LECTURES ON FLUCTUATIONS OF LEVY PROCESSES WITH APPLICATIONS, 2nd Ed., Springer, 2006.

[23] Devroye, L., "An Automatic Method for Generating Random Variates with a Given Characteristic Function," *SIAM Journal of Applied Mathematics*, Vol. 46, No., 4, Aug. 1986, 698-719.

[24] Devroye, L., "Algorithms for Generating Discrete Random Variables with a Given Generating Function for a Given Moment Sequence," *SIAM Journal of Scientific Statistical Computing*, Vol. 12, No. 1, Jan. 1991, 107-26.

[25] Field, T., Tough, R., "Diffusion Processes in Electromagnetic Scattering Generating K-Distributed Noise," *Proceedings of the Royal Society of London A*, Vol. 459, 2003, 5212-23.

[26] Field, T., Tough, R., "Stochastic Dynamics of the Scattering Amplitude Generating K-Distributed Noise," *Journal of Mathematical Physics*, Vol. 44, No. 11, November 2003, 2169-93.

[27] Fayard, P., Field, T., "Discrete Models for Scattering Populations," *Journal of Applied Probability*, Vol. 48, 2011, 285-92.

[28] Barndorff-Nielsen, O., Shiryaev, A., CHANGE OF TIME AND CHANGE OF MEASURE, World Scientific, 2nd ed., 2015.

# Appendix

## Brief Primer on Levy Processes[1]

For the convenience of the reader, this appendix provides a brief primer on Levy processes.

Consider a random process $Y(t)$ with $Y(0) = 0$ and break the interval from 0 to 1 into $n$ equal length segments:

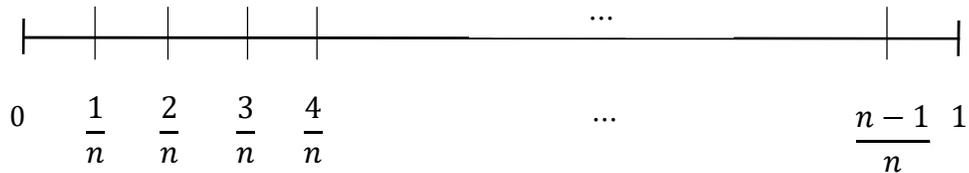

Write

$$Y(1) = Y(1) - Y\left(\frac{n-1}{n}\right) + Y\left(\frac{n-1}{n}\right) - Y\left(\frac{n-2}{n}\right) + \cdots - Y\left(\frac{1}{n}\right) + Y\left(\frac{1}{n}\right) - Y(0)$$

Define

$$y_j^{[n]} = Y\left(\frac{j}{n}\right) - Y\left(\frac{j-1}{n}\right), j = 1, \cdots, n$$

With this notation we have

$$Y(1) = \sum_{j=1}^{n} y_j^{[n]}$$

If we require the random variables $y_j^{[n]}$ to be independent and identically distributed (IID) with distribution $F_n$ – which depends on $n$ – then we have the sum of IID random variables. But, by our construction, the distribution of $Y(1)$ does not change regardless of how we break up the interval [0,1]. The class of random variables for which this is possible is the class of <u>infinitely divisible random variables</u>.

---

[1] This material is largely adapted from [22].

Because the random variables $y_j^{[n]}$ are IID, we may write for the characteristic function of $Y(1)$

$$\phi_{Y(1)}(s) = E[e^{isY(1)}] = \left[\phi_{y_j^{[n]}}(s)\right]^n$$

As an example, consider the case where $Y(1) = X(1)$ is a Gaussian random variable with mean $\mu$ and variance $\sigma^2$:

$$\phi_{X(1)}(s) = e^{i\mu s - \sigma^2 \frac{s^2}{2}}$$

For any positive integer $n$ we may write this as

$$\phi_{X(1)}(s) = \left[e^{i\frac{\mu}{n}s - \frac{\sigma^2 s^2}{n\,2}}\right]^n$$

Thus

$$\phi_{y_j^{[n]}}(s) = e^{i\frac{\mu}{n}s - \frac{\sigma^2 s^2}{n\,2}}, j = 1, 2, \cdots, n$$

This is the characteristic function of a Gaussian random variable with mean $\frac{\mu}{n}$ and variance $\frac{\sigma^2}{n}$.

As a second example, let $Y(1) = P_\lambda(1)$ be a Poisson random variable with intensity $\lambda$:

$$\phi_{P_\lambda(1)}(s) = e^{-\lambda(1-e^{is})}$$

For any positive integer $n$ we may write this as

$$\phi_{P_\lambda(1)}(s) = \left[e^{-\frac{\lambda}{n}(1-e^{is})}\right]^n$$

Thus

$$\phi_{y_j^{[n]}}(s) = e^{-\frac{\lambda}{n}(1-e^{is})}, j = 1, 2, \cdots, n$$

This is the characteristic function of a Poisson random variable with intensity $\frac{\lambda}{n}$.

If we think of each increment $y_j^{[n]}, j = 1, \cdots, n$ as the increment of a process, then we see by the construction above that the process leading to $Y(1)$ has stationary and independent increments. This observation leads naturally to the concept of Levy processes:

> A Levy process is a CADLAG stochastic process $Y(t) = \{Y_t, t \geq 0\}, Y(0) = 0$ with stationary and independent increments. (CADLAG refers to a French expression – *continue a droite, limite a gauche* – and describes a function that is right-continuous and has left limits.)

To explore the notion of Levy processes further, append $m$ versions of $Y(1)$ together:

$$Y(m) = Y(m) - Y(m-1) + Y(m-1) - Y(m-2) + \cdots + Y(1) - Y(0)$$

By the assumption of stationary and independent increments, the characteristic function of $Y(m)$ is given by

$$\phi_{Y(m)}(s) = \left[\phi_{Y(1)}(s)\right]^m$$

Write in general

$$E\left[e^{isY(1)}\right] = e^{-\psi_1(s)}$$

We find

$$\phi_{Y(m)}(s) = e^{-\psi_m(s)} = \left[e^{-\psi_1(s)}\right]^m = e^{-m\psi_1(s)}$$

Hence

$$\psi_m(s) = m\psi_1(s)$$

However we can also break $Y(m)$ in to $n$ segments of length $\frac{m}{n}$ rather than $m$ segments of length 1. In that case we find

$$\psi_m(s) = n\psi_{\frac{m}{n}}(s)$$

The characteristic function of $Y(m)$ is the same in either case; hence we find

$$\psi_{\frac{m}{n}}(s) = \frac{m}{n}\psi_1(s)$$

Because we can represent any $t \geq 0$ arbitrarily closely by appropriate choices of $m$ and $n$ we may now write

$$\psi_t(s) = t\psi_1(s)$$

Thus we have

$$\phi_{Y(t)}(s) = \left[\phi_{Y(1)}(s)\right]^t$$

(Here the CADLAG property, which implies right-continuity, is required to approach $t$ in the limit process.) For our two examples we find

Gaussian: $\phi_{G(t)}(s) = e^{i\mu st - \sigma^2 \frac{s^2}{2}t}$

Poisson: $\phi_{P_\lambda(t)}(s) = e^{-\lambda t(1-e^{is})}$

As a third example consider a compound-Poisson process $\beta(t)$ with intensity $\lambda$ and mixing random variable $X$:

$$\beta(t) = \sum_{i=1}^{P_\lambda(t)} X_i$$

An empty sum is set equal to zero. Here $X_i, i = 1, 2, \cdots$ are IID random variables having distribution $F_X$ and $P_\lambda(t)$ is a Poisson counting process with intensity $\lambda$:

$$Pr[P_\lambda(t) = n] = \frac{(\lambda t)^n}{n!} e^{-\lambda}, n = 0, 1, 2, \cdots$$

$$P_\lambda(0) = 0$$

In this process jumps of random size from the distribution $F_X$ occur at random times determined by a Poisson process with intensity $\lambda$. This process recovers a standard Poisson process if the random variable $X \equiv 1$, i.e. $dF_X(x) = \delta(x-1)dx$. Because $P_\lambda(0) = 0$ it follows that $\beta(0) = 0$. Let $\phi_X(s)$ be the characteristic function of the random variable $X$:

$$\phi_X(s) = \int_{-\infty}^{\infty} e^{isx} dF_X(x)$$

Conditioned on the value of $P_\lambda(t)$ the characteristic function of $\beta(t)$ is given by

$$\phi_{\beta(t)|P_\lambda(t)}(s) = [\phi_X(s)]^{P_\lambda(t)}$$

The unconditional characteristic function becomes

$$\phi_{\beta(t)}(s) = \sum_{n=0}^{\infty} [\phi_X(s)]^n \frac{(\lambda t)^n}{n!} e^{-\lambda t} = e^{\lambda t(\phi_X(s)-1)} = e^{-\lambda t \int_{-\infty}^{\infty}(1-e^{isx})dF_X(x)} = e^{-\lambda t(1-E[e^{isX}])}$$

Define

$$L(x) = \lambda F_X(x)$$

The function $L(x)$ is called the <u>Levy measure</u>. For the compound-Poisson process $F_X$ is the distribution function of a random variable. It then follows that

$$\int_{-\infty}^{\infty} dL(x) = \lambda < \infty$$

This integral is called the mass of the Levy measure $L(x)$.

For another view of the compound-Poisson process, define

$$N(t, A) = \text{number of jumps up to time } t \text{ having size in the set } A$$

For example, let $X$ be a discrete random variable with distribution function given by

$$dF_X(x) = \sum_{k=1}^{\infty} p_k \delta(x - x_k)$$

Thus $X$ takes on the values $x_1, x_2, \cdots$ with corresponding probabilities $p_1, p_2, \cdots$. The value of $\beta(t)$ is equal to the number of jumps (up to time $t$) with size $x_1$ times the value $x_1$ plus the number of jumps with size $x_2$ times the value $x_2$ plus $\cdots$. We therefore can write for this discrete case

$$\beta(t) = \sum_{i=1}^{P_{\lambda_0}(t)} X_i = \sum_{k=1}^{\infty} x_k \, N(t, \{x_k\})$$

For a more general random variable $X$ we then have

$$\beta(t) = \sum_{i=1}^{P_\lambda(t)} X_i = \int_{-\infty}^{\infty} x \, N(t, dx)$$

Here $N(t, dx)$ is the number of jumps (up to time $t$) with size in the interval $(x, x + dx)$.

Examine the characteristic function of $\beta(t)$ with the discrete $X$ discussed above. We find

$$\phi_{J(t)}(s) = e^{-\lambda t \sum_{k=1}^{\infty} p_k(1 - e^{is\,k})} = \prod_{k=1}^{\infty} e^{-\lambda p_k t(1 - e^{isx_k})}$$

This is the characteristic function of the sum of independent Poisson processes with respective intensities equal to $\lambda p_k$ and jump sizes $x_k$. Comparing this result to the result above shows that for fixed $x_k$, $N(t, \{x_k\})$ is a Poisson process with intensity $\lambda p_k$. In the general case we may therefore think of the compound-Poisson process as the sum (integral) of independent Poisson processes with intensity $dL(x) = \lambda dF_X(x)$, where each independent Poisson process has respective jump size $x$. The representation of $\beta(t)$ above therefore shows that for a fixed value of $x$, $N(t, dx)$ is a Poisson process with intensity $dL(x)$ and jump size $x$. Heuristically one may think of $dL(x)$ as the expected number of jumps of height $x$ in an interval of length 1.

As a final example, consider a process comprising the sum of the Gaussian and compound-Poisson processes discussed above, where $G(t)$ and $\beta(t)$ are independent:

$$Y(t) = G(t) + \beta(t)$$

From our results above and the fact that $G(t)$ and $\beta(t)$ are independent we now find for the characteristic function of $Y(t)$:

$$\phi_{Y(t)}(s) = \phi_{G(t)}(s) \phi_{\beta(t)}(s)$$
$$= e^{t\zeta(s)}$$

Here we have defined

$$\zeta(s) = i\mu s - \sigma^2 \frac{s^2}{2} + \int_{-\infty}^{\infty} (e^{isx} - 1)\, dL(x)$$

Heuristically a Levy process is the sum of a constant drift process, a Brownian motion, and a compound-Poisson process. More technically correct, a Levy process can be obtained as the limit of a sequence of such combined processes.

For a compound-Poisson process the Levy mass is finite. However, in general, for a Levy process the mass of the Levy measure need not be finite. The characteristic function of the most general Levy process is obtained by replacing $L(x) = \lambda F_X(x)$ with a more general Levy measure satisfying additional conditions. Provided these other conditions are satisfied, it is possible to have

$$\int_{-\infty}^{\infty} dL(x) = \infty$$

Heuristically this result says that for a general Levy process it is possible to have a countably infinite number of jumps in every interval.

### *Levy Processes with Positive Increments*

In the remainder of this work we limit our discussion to Levy processes for which the characteristic function takes the form

$$\phi_{Y(t)}(s) = e^{t\zeta(s)}$$

$$\zeta(s) = \int_0^{\infty} (e^{isx} - 1)\, dL(x)$$

We also have

$$Y(t) = \int_0^{\infty} x\, N(t, dx)$$

Let $\phi_\tau(s)$ be the characteristic function of a positive random variable $\tau > 0$ and ask if there is a Levy process having increments distributed according to the distribution of $\tau$. The characteristic function of the proposed Levy process $Y(t)$ is given by

$$\phi_{Y(t)}(s) = e^{t\zeta(s)} = [\phi_\tau(s)]^t$$

Therefore we must have

$$\zeta(s) = \int_0^{\infty} (e^{isx} - 1)\, dL(x) = \ln \phi_\tau(s)$$

As an example, we ask if there is a Levy process having a gamma distribution with PDF

$$f_\tau(\tau) = \frac{\nu^\nu}{\Gamma(\nu)} \tau^{\nu-1} e^{-\nu\tau}; \nu > 0$$

The corresponding characteristic function is given by

$$\phi_\tau(s) = \frac{1}{\left(1 - \frac{is}{\nu}\right)^\nu}$$

Thus we must have

$$\int_{-\infty}^{\infty} (e^{isx} - 1) \, dL(x) = -\nu \ln\left(1 - \frac{is}{\nu}\right)$$

A suitable Levy measure $L(x)$ is given by

$$dL(x) = \frac{\nu}{x} e^{-\nu x} dx, x > 0$$

It is important to note, however, that $L$ in this case is not proportional to a CDF. In particular we have

$$\int_0^\infty dL(x) = \int_0^\epsilon dL(x) + \int_\epsilon^\infty dL(x) = \infty$$

The first integral on the right-hand side is infinite for any value of $\epsilon > 0$. The heuristic interpretation of this particular Levy measure says there are an infinite number of "small" (i.e. approaching zero) jumps and a finite number of "large" (i.e. larger than zero) jumps in every interval. Because the Levy mass is infinite, this process cannot be realized as a compound-Poisson process.

The Levy measure of the most general Levy process with non-negative increments satisfies the following condition:

$$\int_0^\infty \min(1, x) \, dL(x) = \int_0^1 x \, dL(x) + \int_1^\infty dL(x) < \infty$$

Note that this condition does not prevent the Levy mass from being infinite. When $\int_0^\infty dL(x) < \infty$, the corresponding Levy process is called a <u>finite activity</u> process. A finite activity process can have zero increments and can be formulated as a compound-Poisson process. On the other hand, when $\int_0^\infty dL(x) = \infty$ the corresponding Levy process is called an <u>infinite activity</u> process. Almost all paths of an infinite activity process have an infinite number of jumps in any compact interval. As a result an infinite activity process cannot have a zero increment and cannot be formulated as a compound-Poisson process. The gamma Levy process discussed above is an example of an infinite activity process.